\documentclass[12pt,preprint,english]{aastex}
\usepackage{graphicx}
\usepackage{amssymb}
\usepackage{times}

\makeatletter

\slugcomment{}
\shorttitle{Coronal Structure \& Stray Light}
\shortauthors{}

\usepackage{babel}
\makeatother

\begin{document}

\title{Solar Coronal Structures and Stray Light in \emph{TRACE}}

\author{C.E. DeForest({*}), P.C.H. Martens(+), and M.J. Wills-Davey(+)}

\affil{({*})Southwest Research Insitute, 1050 Walnut Street Suite 300, Boulder,
CO 80302 USA, deforest@boulder.swri.edu}

\affil{(+)Smithsonian Astrophysical Observatory}

\affil{}

\emph{\hfill\large{(In press, Astrophysical Journal, fall 2008)}\hfill}

\begin{abstract}
Using the 2004 Venus transit of the Sun to constrain a semi-empirical
point-spread function for the \emph{TRACE} EUV solar telescope, we
have measured the effect of stray light in that telescope. We find
that 43\% of 171\AA\ EUV light that enters \emph{TRACE }is scattered,
either through diffraction off the entrance filter grid or through
other nonspecular effects. We carry this result forward, via known-PSF
deconvolution of \emph{TRACE }images, to identify its effect on analysis
of \emph{TRACE }data. Known-PSF deconvolution by this derived PSF
greatly reduces the effect of visible haze in the \emph{TRACE }171
\AA\ images, enhances bright features, and reveals that the smooth background
component of the corona is considerably less bright (and hence more
rarefied) than might otherwise be supposed. Deconvolution reveals
that some prior conclusions about the Sun appear to have been based
on stray light in the images. In particular, the diffuse background
{}``quiet corona'' becomes consistent with hydrostatic support of
the coronal plasma; feature contrast is greatly increased, possibly
affecting derived parameters such as the form of the coronal heating
function; and essentially all existing differential emission measure
studies of small features appear to be affected by contamination from
nearby features. We speculate on further implications of stray light
for interpretation of EUV images from \emph{TRACE }and similar instruments,
and advocate deconvolution as a standard tool for image analysis with
future instruments such as SDO/AIA.
\end{abstract}


\section{Introduction}

All telescopes, including \emph{TRACE } \citep{Handy1998}, scatter
light. The principal scattering mechanisms in a space-based telescope
include diffraction through the aperture and any obscuration in the
beam of the telescope, irregularities or dust on the mirrors themselves,
and reflection or scattering in the detector at the focal plane. All
these effects contribute to forming broad, shallow wings on the point-spread
function (PSF) of the instrument, which describes the image produced
by that instrument when viewing an ideal point source of light. These
wings are typically 3-5 orders of magnitude fainter than the core
of the PSF, but 3-4 orders of magnitude larger, so that a significant
fraction of the light incident into the telescope is spread over a
large portion of the image. 

For discrete scenes such as starfields, broad PSF wings are not greatly
important except that they reduce the net efficiency of the telescope
by reducing the apparent brightness of point sources. For continuous
or near-continuous, high-contrast scenes such as viewed by solar telescopes,
broad scattering wings are quite important: scattering from large
distributed bright structures can overwhelm the emission from dark
regions in the same image. These effects, while well known, have received
little attention from the solar data analysis community when applied
to data from normal-incidence EUV telescopes such as \emph{TRACE,
}but they are present nonetheless and must be accounted for when interpreting
images from \emph{TRACE }(and all other solar telescopes). Not only
quantitative analysis, but even some qualitative interpretations of
\emph{TRACE }data may be compromised if stray light effects are not
accounted for. 

X rays and ultraviolet light are particularly susceptible to scattering
and defocus, and efforts have been made to account for point-spread
function effects in previous instruments. For example, \cite{Maute1981}
attempted blind iterative deconvolution on the X-ray data from \emph{Skylab,}
\cite{svestka1983} applied it to the \emph{SMM/}HXIS instrument;
and \cite{Martens1995} determined a spatially variable PSF for the
\emph{Yohkoh}/SXT instrument. These studies have largely focused on
iterative methods to identify and remove blurring effects caused by
a broad PSF core in the subject instruments, though stray light has
also been an object of study. Stray light deconvolution was commonly
used on SXT data in the later years of that mission (e.g. \citet{Foley1997,Gburek2002,Schrijver2004}). 

The \emph{TRACE} EUV PSF has been studied by several groups. \cite{Lin2001}
used compact, bright flares to study diffraction patterns on the \emph{TRACE}
focal plane and concluded that diffraction from the aluminum filter
grids used in \emph{TRACE} scatters 19\% of incident EUV photons into
a highly structured, broad diffraction pattern; they speculated that
the scattering may be affecting imaging performance. \cite{gburek2006}
used that scattering pattern both to derive both a best-fit PSF core
for \emph{TRACE} and also to determine a portion of the emitted EUV
spectrum from particular flare events. 

In this report, we consider primarily the diffuse scattering wings
of the \emph{TRACE} EUV PSF, and particularly their implications for
interpreting coronal images. The wings are not readily measured using
a point source such as a flare, because the local intensity of the
PSF is quite small far from the core. Outside of diffraction maxima
the weak scattered signal is overwhelmed by local emission even for
bright events such as flares. Deriving the PSF thus requires analysis
of occulted images, using an obstructing body such as the Moon or
a planet. We examined \emph{TRACE }data collected near the times of
several solar eclipses, but did not find a suitable EUV image set
that contained a clear image of the lunar limb. On 2004 Jun 08, Venus
passed in front of the Sun, and several 171 \AA\ image sequences were
collected as the planet traversed the disk of the Sun and the off-limb
corona. We have used those images to derive a semi-empirical scattering
PSF for the \emph{TRACE} 171 \AA\ channel, and have tested the PSF for
correctness by using it to deconvolve several representative\emph{
}images of interesting coronal structures.

In \S\ref{sec:Review-of-Deconvolution} we briefly review deconvolution
and how it is performed, in \S\ref{sec:Constraint-of-the} we describe
the forward modeling process and present our measured PSF, and in
\S\ref{sec:Deconvolution-of-sample} we demonstrate deconvolution
of some representative images. Finally, in \S\ref{sec:Discussion}
we discuss implications for interpretation of EUV coronal images and
recommend deconvolution as a standard reduction pipeline component
for future telescopes.

\section{\label{sec:Review-of-Deconvolution}Review of Deconvolution}

Compensating for the effect of scatter within a telescope requires
\emph{deconvolution}: the telescope convolves the scene with the instrument's
PSF; the effects of the PSF can then be removed by post-processing.\emph{
}Here we briefly review known-PSF deconvolution and how it is performed;
the process is much simpler and more robust than {}``blind deconvolution'',
which does not require a PSF that is known in advance. 

Telescopes in general respond to a point source of light by generating
an image that has finite extent. This image is the PSF of the telescope,
and generally varies at most slowly across the image plane; in this
treatment, we consider it to be constant with respect to position
on the image plane. Images from the telescope are best described as
the convolution\emph{ }of the scene being viewed, with the PSF of
the telescope. The convolution operation spreads out features by integration
(summing) over portions of the source scene, weighted by the PSF.
For simplicity, we consider only the post-sampling image plane and
use discrete operations such as summing, rather than smooth operations
such as integration. 

Convolving an $n_{x}\times n_{y}$ pixel image $I$ with a convolution
kernel $K$ (the instrument's PSF) involves taking a weighted sum
at each location (i,j) in the source image:

\begin{equation}
\left(I\bigotimes K\right)_{i,j}\equiv\sum_{k=-n_{x}}^{n_{y}}\sum_{l=-n_{y}}^{n_{y}}I_{i-k,j-l}K_{k,l}.\label{eq:convolution}\end{equation}
By construction, it is clear that convolution is a linear operation,
so it can be represented with matrix multiplication of $I$ (treated
as a $n_{x}n_{y}$-dimensional column vector) by a $n_{x}n_{y}\times n_{x}n_{y}$
matrix $M_{K}$. Undoing the convolution simply requires inverting
$M_{K}$.

In general, matrix inversion of large matrices is a hard problem (e.g.
\citet{ClaerboutBook}). Fortunately, convolution in real space
is equivalent to elementwise multiplication in Fourier space; in other
words, the Fourier basis diagonalizes $M_{K}$, so that finding its
inverse is trivial. This is the well-known \emph{convolution theorem};
a nice treatment and proof may be found in \cite{Bracewell1999}.
\begin{equation}
I\bigotimes K=\digamma^{-1}\left(\mathcal{I}\cdot\mathbb{\mathcal{K}}\right)\label{eq:conv-theorem}\end{equation}
 where $\digamma$ denotes Fourier transformation, the dot product
represents elementwise multiplication, and curly vectors $\mathcal{I}$
and $\mathcal{K}$ are the Fourier doubles of their italic counterparts
$I$ and $K$. The Fourier transform $\mathcal{K}$ of the PSF is
the \emph{optical transfer function }of the telescope, and its magnitude
$\mathcal{\left|K\right|}$ is the \emph{modulation transfer function}.
Inverting the convolution operation, then, just requires multiplying
by the reciprocal of the optical transfer function:

\begin{equation}
I=\digamma^{-1}(\mathit{\mathcal{I}})=\digamma^{-1}\left(\mathit{\mathit{\mathcal{I\cdot K\cdot}}\mathfrak{R}(\mathit{\mathcal{K}})}\right)\label{eq:inversion}\end{equation}
where $\mathfrak{R}$ is the elementwise reciprocal operator. 

Because Fourier transformation is itself a linear operation, the components
inside the Fourier transform in Eq. \ref{eq:conv-theorem} can be
pulled out, to write:\begin{equation}
I=(I\otimes K)\otimes K^{inv}\label{eq:separation}\end{equation}
where $I\otimes K$ is a source image, and $K^{inv}$ is the function
whose Fourier transform is $\mathfrak{R}(\mathcal{K})$. Eq. \ref{eq:separation}
is useful because it shows that the entire deconvolution operation
can be represented as a convolution by a single inverse PSF. If the
optical transfer function is known, then $K^{inv}$ is trivial to
find. It is just:

\begin{equation}
K^{inv}=\digamma^{-1}\left(\mathfrak{R}\left(\mathcal{K}\right)\right)\label{eq:finding-kinv}\end{equation}
Because of the reciprocal operation, $K^{inv}$ exists only for kernels
with no zero Fourier coefficients. 

Even when the PSF $K$ is known, deconvolution is not quite as simple
in practice as Eq. \ref{eq:finding-kinv} suggests, because of the
presence of noise (which is generally a combination of additive uncorrelated
noise, multiplicative uncorrelated noise, and calibration error).
A typical image can be represented as a (convolved) true focal-plane
image, plus a noise image (which is not wholly independent of the
true image): \begin{equation}
I=I_{true}\bigotimes K_{}+N\label{eq:noise combination}\end{equation}
where $N$ is the noise image. Dividing by $\mathcal{K}_{}$ in the
Fourier plane deconvolves the image but also increases the noise term: 

\begin{eqnarray}
I_{deconv.} & = & \digamma^{-1}\left(I\cdot\mathfrak{R}\left(\mathcal{K}\right)\right)\label{eq:noise-amplification}\\
 & = & I_{true}+\digamma^{-1}\left(\mathcal{N\cdot}\mathfrak{R}\left(\mathcal{K}_{}\right)\right).\label{eq:noise-amp-2}\end{eqnarray}
The noise term is left in Fourier space to emphasize that $\mathfrak{R}(\mathcal{K})$
is a multiplier on the noise level. Most real telescope PSFs attenuate
high spatial frequencies; thus, $\mathfrak{R}(\mathcal{K})$ amplifies
those frequencies in the noise by the same factor. The amplified noise
term at the far right of Eq. \ref{eq:noise-amp-2} can easily overwhelm
$I_{true}$. 

\begin{figure}[!tbh]
\includegraphics[width=6in]{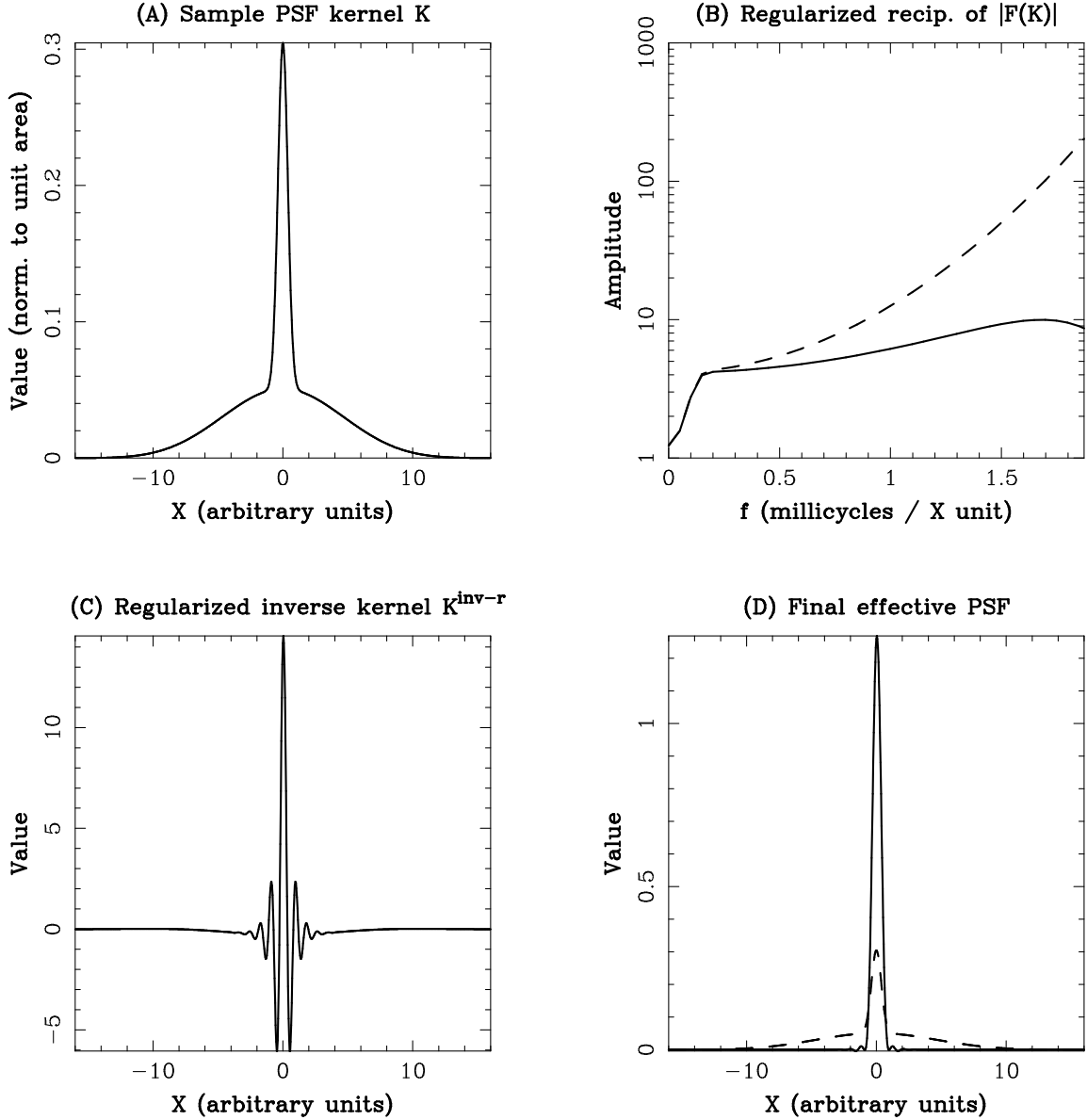}

\caption{\emph{\label{fig:Inversion}Inversion of a sample circularly symmetric kernel
using the Fourier transform and regularized reciprocation. (A): original
sample PSF, $K$; (B): true (dashed) and regularized (solid) reciprocal
of the Fourier components of $K$; (C): resulting inverse kernel $K^{inv}$;
(D): convolution of $K\bigotimes K^{inv}$ would be a delta function
in the ideal case; it is much improved over the original in (A), which
is overplotted as a dashed line.}}

\end{figure}

The theory of \emph{Wiener filters }develops the optimal balance between
signal amplification and noise suppression for additive noise for
linear filters such as deconvolution (e.g. \citet{Press1989}).\emph{
}Rather than develop the ideal Wiener filter for each image, it is
convenient to prevent excessive noise amplification with a \emph{normalized
reciprocal} that rolls over after a certain level of amplification.
We used a simple approximation $\mathfrak{R}_{\alpha,\epsilon}$:\begin{equation}
\mathfrak{R_{\alpha,\epsilon}}\left(z\right)=\left(\frac{z*}{\left|z\right|}\right)\frac{\left|z\right|^{\alpha}}{\left|z\right|^{\alpha+1}+\epsilon^{\alpha+1}},\label{eq:normalized-recip}\end{equation}
 where $z$ is complex but $\alpha$ and $\epsilon$ are real. $\mathfrak{R}_{\alpha,\epsilon}$
converges to $z^{-1}$ for large values of $z$ and to $z*\left|z\right|^{\alpha-1}\epsilon^{-\alpha-1}$
for small values of $\left|z\right|$ (when compared with $\epsilon$),
and rises to a maximum value close to $\alpha\epsilon^{-1}$. Figure
\ref{fig:Inversion} demonstrates construction of an inverse kernel
using $\mathfrak{R}_{0.5,0.05}$ for the reciprocal. 

\section{\label{sec:Constraint-of-the}Constraint of the \emph{TRACE} PSF }

We generated a PSF using data from the Venus transit of 2004 June
08, assuming that Venus emits no EUV light. The images were prepared,
cleaned, and aligned, and a forward model of the scattering was made
by convolving the EUV solar images with a PSF to determine the effect
of the emission on the center of the Venus image. We generated a forward
model PSF that had a narrow core (because we were not interested in
sharpening the images, only in reducing stray light) and included
\emph{a priori} the known 171 \AA\ diffraction pattern first described
by \cite{Lin2001}. Added to the core and the diffraction pattern
was a broad truncated Lorentzian described by three parameters: the
height, the width parameter, and the width of a Gaussian envelope
that was used to truncate the Lorentzian. We convolve the parameterized
PSF with the solar EUV images (with the portion inside the disk of
Venus masked to black) and calculated a model intensity at each of
twelve test loci within the disk of Venus: the center of each of six
images of the planet, and two offset loci in each of three on-disk
images. We compared these model intensities to the original image
brightnesses at the test loci, and adjusted the parameters to find
the best fit for all twelve loci.

\subsection{Image preparation}

\begin{figure}[!htb]
\includegraphics[width=6in]{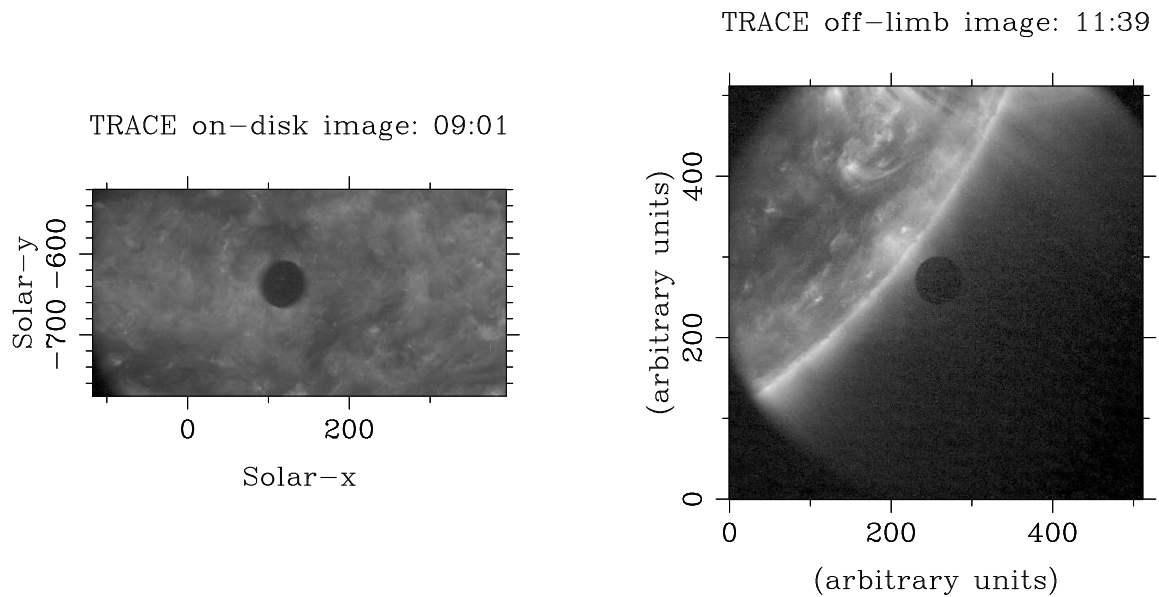}

\caption{\emph{\emph{TRACE} EUV images of the 2004 Venus transit (left) and its immediate
aftermath (right), showing the restricted size of the on-disk field
of view. The images were extended by combining them with the closest
EIT image at the same wavelength. \label{fig:TRACE raw images}}}

\end{figure}

\begin{figure}[!tb]
\center{\includegraphics[height=6.5in]{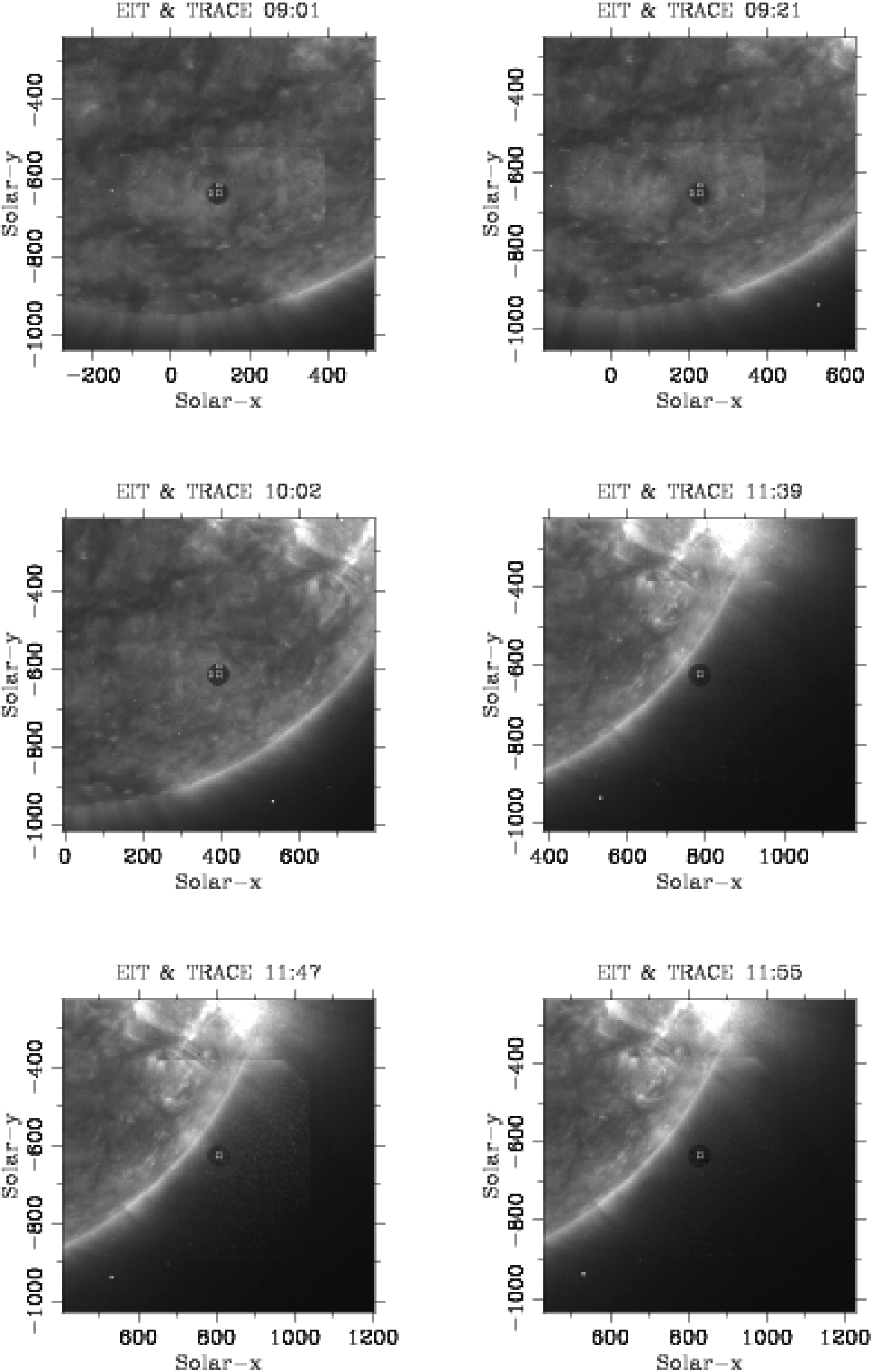}}

\caption{\emph{\label{fig:Venus-regularize}Regularized images from the 2004 June
08 Venus transit used for fitting the \emph{TRACE} scattering PSF.
The EIT image from 19:00 has been used to fill in the missing regions
outside the original field of view. The field of view is 800'' across;
the best-fit PSF has a scattering full-width of \textasciitilde{}300''.
The sample regions are marked: each image is sampled in a 12''x12''
square at the center of the (58'' diameter) disk of Venus, and disk
images are also sampled in a 12''x6'' rectangle offset up 18'' from
the center of the disk, and 6''x12'' rectangle offset left 18'' from
the center of the disk.}}

\end{figure}

Two sets of \emph{TRACE} EUV images are present from the Venus occultation,
both in the 171 \AA\ passband. During the transit itself many images
were taken with no binning and either 16 s, 30 s, or 90 s exposure
(in the hour 09:00 - 10:00 UT); and shortly after the transit a series
of images with 16 second exposure and 2x2 binning were collected off-limb
(in the hour 11:30 - 12:30 UT). On-disk, we used the 90 s exposures
to minimize background noise; off-limb, we median-filtered blocks
of 5 images along the time axis, to reduce background noise in those
less-well-exposed frames. Figure \ref{fig:TRACE raw images} shows
a sample on-disk and off-limb image. We downloaded Level 1 data directly
from the \emph{TRACE} web site. We further corrected the zero point
by subtracting the average of the 30x30 pixel region in the lower-left
corner of each \emph{TRACE} image (in the filter vignetted area),
then scaled the images to 1'' per pixel (thus binning 2x2 the on-disk
images) and divided out the exposure time and binned pixel size to
arrive at calibrated images in \emph{TRACE} digitizer numbers (DN)
arcsec$^{-2}$ s$^{-1}$, so that the images were directly comparable.
We despiked each image using a simple unsharp-mask + threshold algorithm
(the {}``spikejones.pdl'' routine in the PDL portion of the \emph{Solarsoft}
(\citet{FreelandHandy1998}) software distribution), and replaced
each spike value with the median of valid values in its neighborhood
in the same image. We made a 5-image pixelwise median of the off-limb
images, then spatially shifted each image to center Venus in the frame
(Figure \ref{fig:Venus-regularize}). 

Because \emph{TRACE} has a limited field of view, but the primary
and secondary mirrors are exposed to the entire solar disk, it was
necessary to extend the field of view to a good fraction of the solar
disk to model the extended PSF and reduce the possibility that edge
effects would affect the result. We used the closest-in-time full-disk
171\AA\ image from \emph{SOHO}/EIT (\citet{Delaboudiniere1995}), collected
at approximately 19:00 UT on the same day, to fill in missing values
outside the \emph{TRACE} field of view. Although the EIT image was
collected some 7-9 hours after the Venus occultation data, the portions
of the solar image that are affected are far from test loci in the
disk of Venus, and therefore only large spatial scales are important;
brightness on these scales varies on timescales of hours to days.
The EIT image was prepared using the instrument-supplied eit\_prep
software, scaled to 1'' per pixel, derotated to the \emph{TRACE} time,
and multiplied so that a 100x100 pixel sum (chosen to be far from
the \emph{TRACE} vignetted regions) was equal between the EIT and
\emph{TRACE} image. Then the dark (vignetted) portions of the \emph{TRACE}
image were replaced with the corresponding portion of the scaled,
corrected EIT image. Figure \ref{fig:Venus-regularize} shows all
of the resulting cleaned, combined images and the loci within them
that were used to constrain the fit.

\subsection{Forward Modeling of the PSF}

We forward modeled the scattering portion of the \emph{TRACE} PSF
as the sum of the measured diffraction pattern \citep{Lin2001,gburek2006}
and a circularly symmetric scattering profile produced by revolving
a radial function about the origin. The revolved function was a sum
of a narrow core, a {}``shoulder'' Gaussian, and a truncated Lorentzian
intended to represent the scattering wings. The central core width
was chosen to have a value much less than the Gburek et al. (2006)
width, because the intent is to remove scattering wings rather than
to sharpen the core of the PSF by deconvolution. The diffraction pattern
was convolved with the central core to avoid pixelization artifacts
due to the delta functions in it.

The analytic formula is:\begin{equation}
K_{\alpha,w_{},\sigma}(r,\theta)=\gamma^{-1}\left(\left(e^{-4ln(2)r^{2}}/1.27\right)\otimes D(r,\theta)+\frac{\alpha}{r^{2}/w^{2}+1}e^{-4ln(2)r^{2}/\sigma_{t}^{2}}+\beta e^{-4ln(2)r^{2}/\sigma_{s}^{2}}\right),\label{eq:PSF}\end{equation}
where $r$ is distance in the image plane, measured in arcseconds;
the 1.27 normalizes the integral under the first Gaussian to unity;
$\otimes$ represents convolution; $D(r,\theta)$ is the diffraction
pattern described below, including a central core; $\alpha$ is relative
strength of the Lorentzian wings; $w$ is the width of the Lorentzian;
$\sigma_{t}$ is the FWHM of the truncating Gaussian (in arcsec);
$\beta$ is the strength of a Gaussian shoulder to the curve; $\sigma_{s}$
is the width of the shoulder; and $\gamma$ is a factor to normalize
the integral under the 2-D convolution kernel to unity. 

The parameters were found by an iterative fit method: for each guess
set of parameters, the kernel was calculated on an 800''x800'' grid
at 0.5'' resolution, multiplied by each of the six composite images,
and summed to find the expected scattered intensity at the center
of each Venus image. Note that the \emph{TRACE} field of view is only
about 500'' across -- the larger FOV (available with the EIT overlay)
was used so that the \emph{TRACE} field of view, itself, wouldn't
constrain the fits. 

In addition, we used an offset kernel to calculate the intensity at
an off-center locus in each of the three on- disk images, to constrain
the shoulders of the curve a few arcseconds from the core. We did
not use the off-center brightness in the off-limb Venus images, because
pixels above the limb of the Sun probably contain proportionally more
scattered light than do pixels on the solar disk. The geometry of
each sample point is shown in Figure \ref{fig:Venus-regularize}.
We compared the intensities to the forward scattering model, and adjusted
the parameters initially {}``by eye'' to find a reasonable match
with the twelve data points. Finally, we optimized the fit with an
amoeba algorithm (e.g. \citet{Pressbook}), holding $w_{core}$
at the conservative 0.5 $arcsec$ full-width at half-maximum and penalizing
errors in the overcompensation direction (taking the image value below
zero) a factor of 100x worse than errors in the undercompensation
direction. The resulting parameters are given by:

\begin{equation}
\begin{array}{ccc}
\alpha & = & 2.06\times10^{-5}\\
w & = & 57.7\, arcsec\\
\sigma_{t} & = & 68.4\, arcsec\\
\beta & = & 6\times10^{-7}\\
\sigma_{s} & = & 15\, arcsec\end{array}\label{eq:best-fit-params}\end{equation}

\begin{figure}[!htb]
\center{\includegraphics[bb=0bp 0bp 240bp 226bp,clip,width=4in]{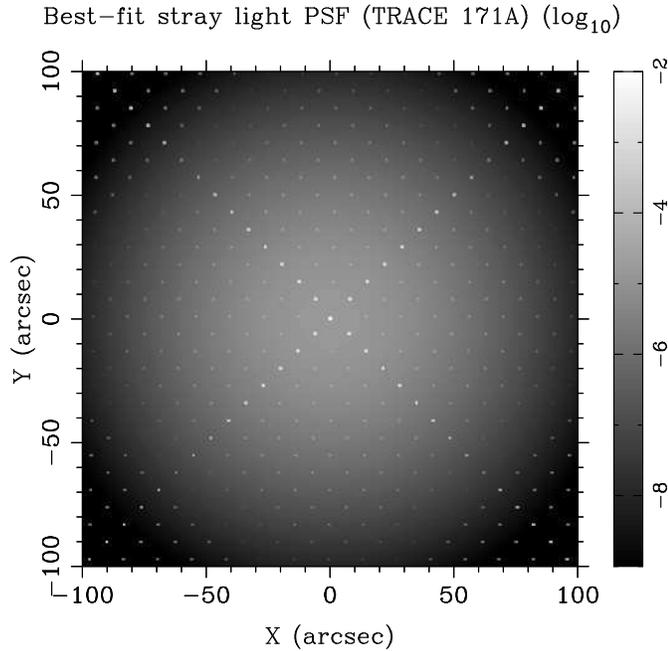}}

\caption{\emph{\label{fig:PSF}The best-fit point-spread function (and its inverse)
for the 171\AA\ channel of \emph{TRACE}. The intensity is plotted on
a log scale and is normalized to an integral of unity. }}

\end{figure}

The fit is within $0.021\, ct\, arcsec^{-2}\, s^{-1}$ of the measured
data value at each sample point and 0.019 $ct\, arcsec^{-2}\, s^{-1}$
RMS across the 12 data points, compared to absolute brightnesses of
0.5-1 $ct\, arcsec^{-2}\, s^{-1}$ in the interior of the disk of
Venus. The fitted PSF and its inverse are plotted in Figure \ref{fig:PSF}.
There is no significant contribution to the total energy outside of
a 100'' radius. The maximum intensity in the PSF core is 0.5; hence,
the modeled isotropic scattering function is down by 5 orders of magnitude
from the center of the PSF. Nevertheless, its large cover compared
to the core of the PSF yields a significant amount of scattering.

\begin{figure}[!htb]
\center{\includegraphics[width=3in]{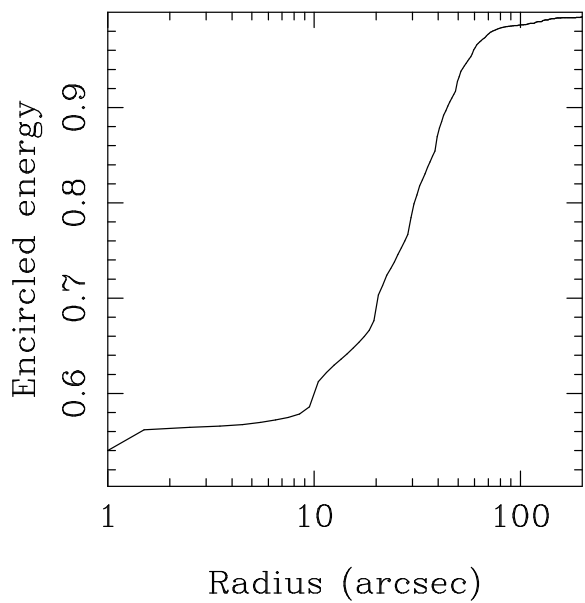}}

\caption{\emph{\label{fig:Encircled-energy}Encircled energy vs. distance for the
model PSF calculated with Equations \ref{eq:PSF}-\ref{eq:best-fit-params}.
About 57\% of the PSF's total energy is contained in the central core,
and 43\% is contained in broad scattering wings that extend about
100'' (200 \emph{TRACE} pixels) in all directions. The {}``wobbles''
in the curve are due to successive diffraction maxima.}}
\end{figure}

The encircled energy is plotted versus distance in Figure \ref{fig:Encircled-energy}.
The encircled energy curve can be counter-intuitive at first: while
the scattering wings start and remain small (at under $10^{-3}$ of
the intensity at the core of the PSF), at each successive radius more
area is available to contribute to the total integrated energy. Hence,
in a nearly uniform scene most of the stray light at a given point
in the image plane arises from features 20-50 $arcsec$ (40-100 \emph{TRACE
}pixels) away. Approximately 43\% of the energy in the derived PSF
exists more than 2'' from the center of the core.

\section{\label{sec:Deconvolution-of-sample}Deconvolution of sample \emph{TRACE}
images}

\begin{figure}[!tb]
\includegraphics[width=6in]{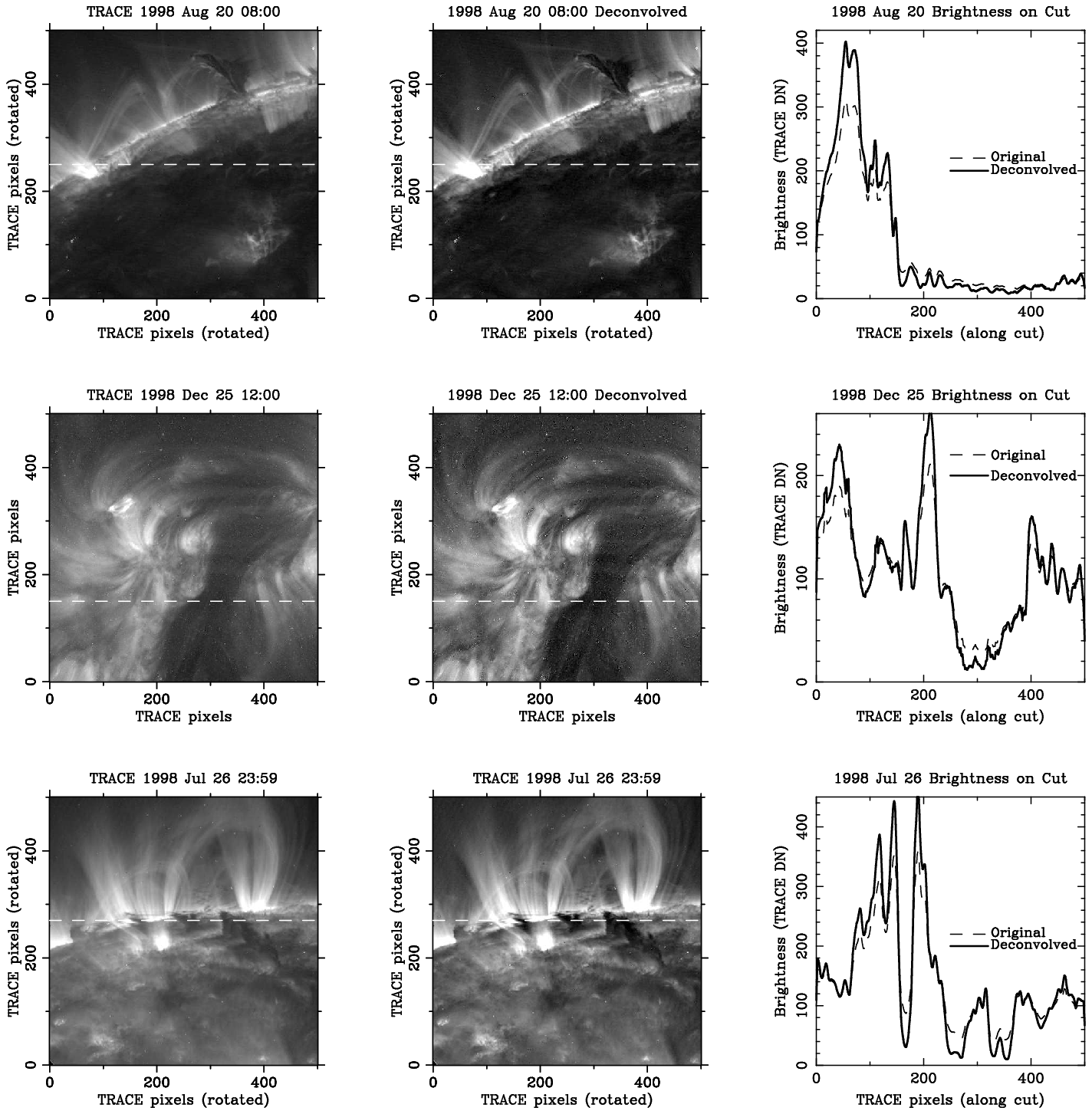}

\caption{\emph{\label{fig:deconvolution-demo}Deconvolution of sample \emph{TRACE}
171 images greatly increases contrast of dark features, eliminates
`haze'. Top: limb scene. Center: disk AR. Bottom: Prominences near
the limb. In each row, left is a level-1 processed \emph{TRACE }image;
middle is the same image, deconvolved; and right is a plot of brightness
along the indicated cut. Compact bright features (such as the small
active region at the left hand side of the top panel) are increased
by nearly 50\% in brightness; dark features in the midst of bright
regions (such as the dark lane near the center of the central image,
or the dark prominences at bottom) are darkened by a factor of about
3. Note that the prominence at far left of the lower panel is behind
a large bright loop structure, while the others in that image are
not.}}

\end{figure}

Scattering of 43\% has a significant impact on images collected with
\emph{TRACE.} Here we present results of deconvolution, using the
measured PSF from \S\ref{sec:Constraint-of-the}. A broad variety
of 171\AA\ images were deconvolved and tested for correctness. In no
case did any of the pedestal-subtracted, deconvolved images have significant
negative-flux regions, an indication that the fitted PSF is either
correct or conservative compared to the real PSF of the instrument.
Isolated pixels may be carried below zero, due to JPEG artifacts or
photon counting noise, but smoothing the image with a $5\times5$
pixel boxcar kernel eliminates the negative regions. In general, bright
features get marginally brighter, and dark regions get much darker,
after deconvolution. Figure \ref{fig:deconvolution-demo} shows the
results of deconvolution of a limb scene, a near-disk-center scene,
and several dark prominences near the limb. All of the images have
been cropped to the middle $500\times500$ pixels of the \emph{TRACE
}detector after deconvolution, to avoid edge effects.

The images were despiked using a spatial spike finder
({}``spikejones.pdl'' in the \emph{Solarsoft }distribution; \citet{FreelandHandy1998})\emph{
}In each case, the median value of the lower, left 15x15 pixel region
from each despiked \emph{TRACE} image was used as a zero-point reference:
because \emph{TRACE }is vignetted by the thin-foil filter ring on
board, the lower-left corner serves as a reasonable dark reference
value. No additional background subtraction was performed.

In general, contrast is greatly enhanced throughout the images. For
example, the lane in the disk-center active region (center row of
Figure \ref{fig:deconvolution-demo}) is shown to be about a factor
of 3 darker than might be expected from na\"ive analysis of the image
(without deconvolution), and small features embedded in bright regions
(such as the fan of threads on the right hand side of that image)
can be as much as doubled in contrast relative to their local background.
The prominences in the bottom row of Figure \ref{fig:deconvolution-demo}
demonstrate the effectiveness of deconvolution at removing nearby
coronal brightness: the prominences (which protrude about one density
scale height at 1MK, hence two intensity scale heights) are shown
to be quite dark, presumably because they protrude above most of the
quiet coronal emissions. The leftmost prominence is directly behind
a loop base and is therefore not darkened nearly as much as the others;
this forms a good check that the deconvolution is not simply darkening
features arbitrarily. The second-from-left prominence is seen between
two bright loop structures, and is hence quite dark despite the apparent
bright foreground.

\section{\label{sec:Discussion}Discussion}

The result that \emph{TRACE} images contain a significant amount of
scattered light is not, in itself, new. Most telescope PSFs include
scattering wings, and \emph{TRACE }is no exception. The filter grid
in the front of the telescope is known to scatter $\sim20\%$ of incident
light into a structured diffraction pattern with myriad local maxima
(\cite{Lin2001,gburek2006}). The present analysis is new in three
important respects: it is (to our knowledge) the first analysis of
scattered light using \emph{TRACE }occultation data to derive a PSF;
much more scattered light is found than can be accounted for merely
by diffraction; and the process is taken to its natural conclusion
of deconvolving the original \emph{TRACE }images to show the effect
of the stray light on scientific interpretation of the images. 

The \emph{TRACE }Venus data are not detailed enough to constrain a
highly structured PSF model, but the simple empirical fit described
here is sufficient to improve existing images via deconvolution, and
passes the most basic of deconvolution tests, suggesting that it is
not overcompensating for the scattering wings. Deconvolution with
our scattering PSF has a similar effect to background subtraction
on the interpretation of small features: for features small compared
to the scattering wings, the effect of the surrounding bright features
is approximately constant, so subtraction of a modeled or fixed background
yields similar effects in particular local areas of a given \emph{TRACE
}image. The principal advantages of deconvolution are that an approximation
of absolute brightness is reproduced, rather than the offset relative
brightness that may be extracted from simple background-subtracted
images; and that moderate-scale features are treated correctly (they
are not treated correctly by simple background or pedestal subtraction). 

Deconvolution not only darkens the faintest portions of the image,
it both increases the relative contrast of small bright features embedded
in a bright background and also affects photometric estimates of the
relative density of any small bright feature seen with \emph{TRACE}.
This describes several features of interest in the \emph{TRACE} data,
including active-region threads that are a subject of current debate
\citep{WatkoKlimchuk2000,WarrenWinebarger2003,Fuentes2006,DeForest2007}. 

The greater coronal contrast we find in deconvolved \emph{TRACE} images
gives indirect support to the idea that the corona is close to hydrostatic
equilibrium despite the observed tallness of bright features such
as active region loops. The coronal density scale height is about
50 Mm at 1 MK, so the emissivity scale height of Fe IX \& Fe X emission
line features (close to 1MK ionization temperature) is about 25 Mm
(0.035 $R_{S}$) assuming local thermal equilibrium. Thus the EUV-visible
corona might be expected to form a thin layer near the photosphere
with no significant emission arising at altitudes higher than about
0.07 $R_{S}$. Essentially all \emph{TRACE }EUV images show significant
background brightness high in the corona; the brightness is visible
above the detector {}``pedestal'', because there is contrast between
dark but {}``live'' pixels that are part of the image, and corner
pixels that are vignetted by the round filter mount at the back of
the instrument. The current measured instrument PSF suggests that
most or all of this background brightness is due to scattering within
the telescope, because dark features (such as the prominences in the
bottom row of Figure \ref{fig:deconvolution-demo}) are reduced nearly
to zero brightness when deconvolved. This is the general behavior
to be expected from a thin hydrostatic atmosphere at a particular
temperature: tall features that are more than 1-2 scale heights tall
should have little or no emission above them.

Active region loops appear to have a large scale height compared to
that expected for 1-2 MK plasma \citep{SchrijverMcMullen2000,AschwandenNitta2000,WinebargerWarrenMariska2003,Fuentes2006,DeForest2007}.
Three explanations that have been advanced are resonant scattering
of EUV (which varies as $n_{e}$ rather than $n_{e}^{2}$), support
by non-hydrostatic momentum transport mechanisms such as siphon flows
or wave motion, or geometric considerations that attenuate brightness
at the bases of the loop. Our result that the quiet corona appears
to be consistent with the expected hydrostatic scale height seems
to eliminate resonant scattering as a mechanism for tallness, because
it would imply a stronger haze in the foreground at high solar altitudes.
Further, it seems to limit the functional form of anomalous support
mechanisms that could lengthen active region loops' scale height,
because such mechanisms must act preferentially against active region
loops and not quiet sun loops, to be consistent with the morphology
of the deconvolved images. This can further be construed as circumstantial
evidence for a geometric, rather than intrinsic, explanation for active
region loops' long apparent length \citep{DeForest2007}. 

\cite{Moore2008} have recently used image-processing techniques
to separate the hazy and sharp components of active region loops viewed
with \emph{TRACE. }Such analyses rely on the sharp component of the
corona as an indicator of stray light. With 43\% scattering of the
total light incident on the telescope, 57\% is left to be focused;
hence, we expect that the hazy portion of such separated image pairs
derived from \emph{TRACE }171\AA\  data should contain about 75\%
as much total brightness as the sharp portion does, on the basis of
stray light alone. 

In addition to morphological differences, corrections to the relative
brightness of features such as active region threads and voids affect
parameters such as the derived Alfv\'en speed, because of the $n_{e}^{2}$
dependence of EUV emission. Structures with spiky density profiles
emit more EUV per electron than do smooth structures, and the inferred
Alfv\'en speed depends both on the magnetic field and the derived electron
density. Onset of some \emph{TRACE-}observed EIT waves appears to
require high Alfv\'en speeds of up to $3\, Mm\, s^{-1}$ \citep{willsdavey2007};
this high speed is difficult to explain in the presence of a diffuse
background corona around the source region of the EIT wave. If in
fact active regions contain nearly evacuated regions (as in the center
panel of Figure \ref{fig:deconvolution-demo}), then the variation
in Alfv\'en speed is greatly increased and the region-wide average Alfv\'en
speed may be significantly higher than would otherwise be inferred.

As a final example of the impact of stray light in EUV images\emph{,
}coronal heating properties have been derived \citep{Schrijver2004}
by examining the contrast between coronal holes and bright structures,
and may be affected by scattering in \emph{SOHO}/EIT and/or \emph{Yohkoh}/SXT.
Specifically, if coronal holes are significantly darker, and bright
structures are significantly sharper and brigher, than is apparent
in raw EUV and X-ray images, then the coronal heating mechanism may
not be as distributed as might otherwise be inferred.

The model PSF that we have derived for \emph{TRACE} is somewhat simplified:
we have included, \emph{a priori, }detailed structure that is known
from earlier studies \citep{Lin2001}, and parameterized
an additional scattering term based on the empirical behavior of stray
light on rough mirrors. We have not taken into account possible anisotropy
or spatial variability of the scattering, attempted to gain physical
understanding of the causes of the PSF, or modeled scattering phenomena
that do not fit within the paradigm of a simple PSF. Based on measurements
of the Venus transit in 2004, we have found that roughly 43\% of incident
energy is scattered by \emph{TRACE}, so that approximately half of
the scattered energy may be ascribed to the diffraction pattern found
by Lin et al. and approximately half to other mechanisms. Deconvolution
greatly improves contrast in \emph{TRACE }images, raising concerns
about the interpretation of those images.

More generally, deconvolution to increase contrast in images with
scattering wings is strongly recommended for observation from present
and future EUV and X-ray telescopes. We have shown that deconvolution
can greatly affect the contrast of observed features, and discussed
how this effect may affect a broad variety of science questions. Further,
deconvolution to remove broad scattering wings is in general not as
hazardous to the data as is deconvolution to increase sharpness in
the core of the telescope PSF. That is because there are high spatial
frequencies present in the core of the kernel, even if it is added
to a much broader distribution, so that noise is not increased as
much from deconvolution of a scattering PSF as from a broad PSF core.

\acknowledgements{Thanks to the \emph{TRACE} and \emph{SOHO}/EIT teams for making their
data available to everyone. We also owe thanks to L. Golub, J. Cirtain,
K. Schrijver, and H. Throop for illuminating discussions, and to the
anonymous referee for several suggestions that improved the work.
SOHO is a project of international cooperation between NASA and ESA.
This work was funded under NASA's SHP-SR\&T program.}

\end{document}